\documentclass[preprintnumbers,twocolumn,showpacs,aps,prl,superscriptaddress]{revtex4-1}

\usepackage{lineno}  
\usepackage{graphicx}
\usepackage{dcolumn}
\usepackage{bm}
 \usepackage{amsmath}
 \usepackage{epsfig}

 \input babarsym

 \input Definitions

\begin{document}
\preprint{\hbox to\hsize{BABAR-PUB-18/004\hfil}}
\preprint{\hbox to\hsize{SLAC-PUB-17489\hfil}}
\vspace*{1cm}

\title{Measurements of the Absolute Branching Fractions of \boldmath$B^\pm \to K^\pm X_{c\bar c}$}

\author{J.~P.~Lees}
\author{V.~Poireau}
\author{V.~Tisserand}
\affiliation{Laboratoire d'Annecy-le-Vieux de Physique des Particules (LAPP), Universit\'e de Savoie, CNRS/IN2P3,  F-74941 Annecy-Le-Vieux, France}
\author{E.~Grauges}
\affiliation{Universitat de Barcelona, Facultat de Fisica, Departament ECM, E-08028 Barcelona, Spain }
\author{A.~Palano}
\affiliation{INFN Sezione di Bari and Dipartimento di Fisica, Universit\`a di Bari, I-70126 Bari, Italy }
\author{G.~Eigen}
\affiliation{University of Bergen, Institute of Physics, N-5007 Bergen, Norway }
\author{D.~N.~Brown}
\author{Yu.~G.~Kolomensky}
\affiliation{Lawrence Berkeley National Laboratory and University of California, Berkeley, California 94720, USA }
\author{M.~Fritsch}
\author{H.~Koch}
\author{T.~Schroeder}
\affiliation{Ruhr Universit\"at Bochum, Institut f\"ur Experimentalphysik 1, D-44780 Bochum, Germany }
% ...
\collaboration{The \babar\ Collaboration}
\noaffiliation

\begin{abstract}

A study of the two body decays \bch\to X$_{c\bar c}$\Kch, where X$_{c\bar c}$ refers to one charmonium state,  is reported by the \babar~ collaboration
using a data sample of 424 \invfb. The absolute determination of branching fractions for these decays are significantly improved compared to previous \babar~ measurements. Evidence is found for the decay  \Bp\to $X(3872)$\Kp at the $3\sigma$ level. The absolute branching fraction ${\cal B}(\Bp\to X(3872)\Kp) = (2.1\pm0.6$(stat)$\pm$0.3(syst))$\times10^{-4}$ is measured for the first time. It follows that ${\cal B}(X(3872)\to J/\psi\pi^+\pi^-)=(4.1\pm1.3)$\%, supporting the hypothesis of a molecular component for this resonance.

\end{abstract}


\maketitle

 In  two-body $B$  decays $B \to XK$, the $X$ particle is predominantly a $c\overline{c}$ 
 system with large available 
 phase space. Many charmonium states are thus  produced, with approximately equal rates when no strong selection rules apply~\cite{quigg}.
 They  have mostly been 
 observed using an exclusive reconstruction of the  charmonium state 
 $X_{c\bar c}$ (\etac, \jpsi, $\chi_{c1}$, $\chi_{c2}$, \etacp, $\psi'$), with possibly the associated observation of the decay  \bch\to $X_{c\bar c}$\Kch~ \cite{babar1,belle2}. The exotic charmonium state $X(3872)$, also known as $\chi_{c1}(3872)$, has also been reconstructed in this way \cite{bellex38,babarx38}.
 
The determination of the absolute branching fraction  ${\cal B}$(\Bp\to $X(3872)$\Kp) leads to the absolute ${\cal B}$($X(3872)$\to$J/\psi\pi^+\pi^-$), bringing useful information regarding the complex nature of the $X(3872)$. The original tetraquark model \cite{maiani} predicts this branching fraction to be about 50\%. A more refined tetraquark model \cite{maiani2} can accommodate  a much smaller branching fraction, but requires  another particle, $X(3876)$, not yet observed. Various molecular models \cite{braaten,barnes,ortega} predict this branching fraction to be  $\lsim10$\%. 
Using the  $X(3872)$ total width determination based on its line shape, or an upper limit on this quantity, information is provided on  the partial width $\Gamma(X(3872)$\to$J/\psi\pi^+\pi^-$), for which a wide range of predictions exist, from 1.3 MeV in the case of  a pure charmonium state \cite{swanson}, to about 100 keV for molecular models  \cite{braaten}.

In this study, we adopt a technique,  pioneered by \babar~\cite{charmo2005} and re-used by Belle \cite{belle:similar}, based on the measurement  in the $B$ rest frame of the kaon momentum spectrum, where each
two-body decay is identified by  its  monochromatic kaon. Taking advantage of the $\Upsilon (4S)$  decay to a $B\bar B$ meson pair, the $B$ center-of-mass frame  is determined event-by-event by fully reconstructing the other $B$ meson.  The branching fractions
for the two-body decays \bch\to X$_{c\bar c}$\Kch~ can thus be measured
independently of any {\it a priori} knowledge of the $X_{c\bar c}$ decay properties.

We use a data sample with an integrated luminosity of 424 fb$^{-1}$~\cite{babarlumi},  collected with the 
\babar~detector at the PEP-II storage ring, at a center-of-mass energy corresponding to the
$\Upsilon (4S)$ mass.
Charged tracks are reconstructed with a 5-layer silicon vertex 
tracker (SVT) and a 40-layer drift chamber (DCH), located in a 1.5 T
 magnetic field generated by a superconducting solenoidal magnet. 
The energies of photons and electrons are
measured with a CsI(Tl)  electromagnetic calorimeter (EMC).
Charged hadron identification is performed using ionization measurements 
in the SVT and DCH and using a ring-imaging \v{C}erenkov detector (DIRC).
The instrumented flux return of the solenoid (IFR) is used to identify muons.
A detailed description of the \babar~detector can be found in Ref.~ \cite{babardet1,babardet2}. 

The analysis method is similar to that presented in Ref.~\cite{charmo2005}. The complete reconstruction of one of the two $B$ mesons  provides access to the rest frame of the other $B$ meson. For signal events, two-body $B^\pm$ decays to $K^\pm X$, the kaon momentum in the $B$ center-of-mass frame, $p_k$,  exhibits a peak for each $X$ particle, with mass $m_X=\sqrt{m_B^2 + m_K^2 - 2 E_K m_B}$, where 
$m_B$ and $m_K$ are the masses of the $B$ and $K$  mesons and $E_K$ is the kaon energy in the $B$ rest frame. 
 The $p_k$  spectrum contains, besides a series of signal peaks,  
a background due to kaons from non-two-body decays or 
from decays of charmed mesons. We determine the observed number of each charmonium resonance $X_{c\bar c}$
from a fit to the kaon momentum distribution. 

\begin{figure}
 \epsfig{figure=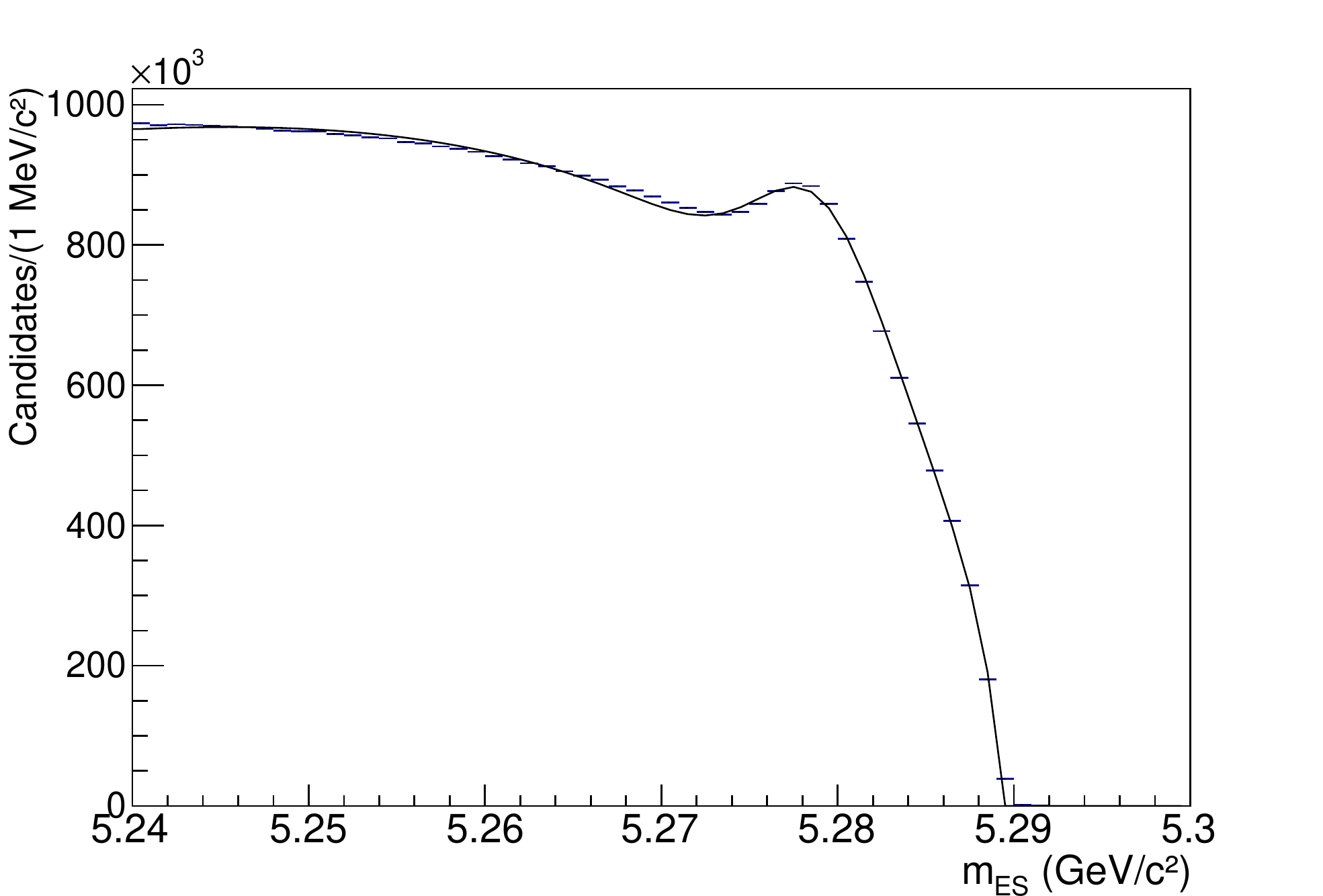,width=.49\textwidth}
\caption{The \mes distribution of the exclusively reconstructed $B^\pm$, with the fit result superimposed.}
 \label{data_bchbch_fig}
 \end{figure}

Event selection requires the reconstruction of a tagging \bch\ meson  ($B$-tag) from $B\to SY$ decays, where the seed $S$ is a fully reconstructed $D^{(*)0}$, $D^{(*)\pm}$, $D_s^{(*)\pm}$, or \jpsi meson, and  
$Y$ represents a combination of  $\pi^\pm$, $K^\pm$, $\pi^0$, and $K_S^0$ hadrons~\cite{fullreco}. For each mode, a purity (defined as $S/(S+B)$, where $S$ is the number of signal events and $B$  the number of background events) larger than  0.08 is required. 
The number of $B$ candidates is determined with a fit, shown in Fig.~\ref{data_bchbch_fig},  to the
distribution of the $B$-energy-substituted mass, $m_{ES} = \sqrt{E_{\rm{CM}}^2/4 - p_B^2}$. Here,
$E_{{\rm CM}}$ is the total center-of-mass energy, determined from the beam parameters, and $p_B$ is the measured momentum of the \
reconstructed $B$ in the $\Upsilon (4S)$ rest frame.
The fit function is the sum of a Crystal Ball function~\cite{crystalball} describing the
signal  and an ARGUS function~\cite{argus} for the background. 
The number of fully reconstructed \bch\  decays found by the fit is $1.65\times 10^6\pm 4\times 10^3{\rm (stat)}\pm 6\times 10^4{\rm (syst)}$. The systematic uncertainty is dominated by the background shape near the kinematic end point. This  event yield is mentioned for reference but is not used in the determination of the branching ratios, except for the cross-check on BR(\bch\to\jpsi\Kch).

 If more than one $B$ candidate is found in an event, all candidates are retained. This is an important difference compared to Refs.~\cite{charmo2005} and  \cite{belle:similar}, where only one candidate per event is retained. This method  increases the efficiency and provides  better  decoupling between the signal and tag sides. Events not considered before, where the candidate selected as the best one was not the correct one, are now retained, including those where it belonged to the signal side. This point is important for the $X(3872)$ measurement because the probability to reconstruct a candidate from the signal side is enhanced for particles decaying to $D$ mesons. The new method provides efficiency gains up to a factor 3.
The mean number of  $B$-tag candidates per event is 1.85 and 39\% of events have more than one candidate.

Event selection criteria are as follows: Each $B$-tag candidate should have $\mes>5.275$ GeV/$c^2$
and be accompanied by an opposite-sign kaon candidate (charge conjugation is always implied), passing a tight particle identification selection. The pion contamination in this kaon sample is below 2\%. A neural network (NN) is then used to suppress the continuum background.
The inputs to the NN are seven variables related to the reconstructed $B$ 
characteristics, to its production kinematics, to the topology of the full 
event, and to the angular correlation between the reconstructed $B$ and the 
rest of the event.
The NN selection has an 80\% efficiency for generic $B^+B^-$ events and a factor 10 rejection against  non-$B$ background events coming from $u, d, s$, or $c$ quark-antiquark pairs.

A second NN is used to reject secondary kaons produced in $B$-daughter $D$ meson decays. This is  a large background that increases rapidly with decreasing kaon momentum. 
In the $B$ rest frame, the secondary kaons  are  embedded  in the $D$ decay products, which, given the boost of the $D$ meson and its mass, are bounded in a cone and form a wide jet, whereas signal-kaons  recoil against a 
massive (3 to 4 GeV/$c^2$) state and tend to be more isolated, with  the rest of the $B$ decay products being more spherical. The input variables to this NN are: the energy contained in a cone around the kaon track, the sphericity of the system recoiling against the kaon, the angle between the kaon and the thrut axis of the recoiling system, the minimum mass formed with the kaon and the recoiling particles~\cite{supplement}.
The two NN are then combined in a single neural net, called SuperNN, to optimize further the signal to background.
Because of the non-negligible variation of the event topology with the mass of the charmonium particle, the SuperNN is trained separately in  the \jpsi and \etac\ signal region, and in the \psip\ and \etacp\ region, with kaon background taken from simulation in the momentum ranges 1.6--1.9 and  1.2--1.5 GeV/$c$, respectively. 
The SuperNN performance corresponds to a  72\% signal efficiency at the $X(3872)$ peak and a  background rejection  factor varying between three in the $X(3872)$ and \psip\ region to 4.5 in the \jpsi region. 

\begin{figure}
\epsfig{figure=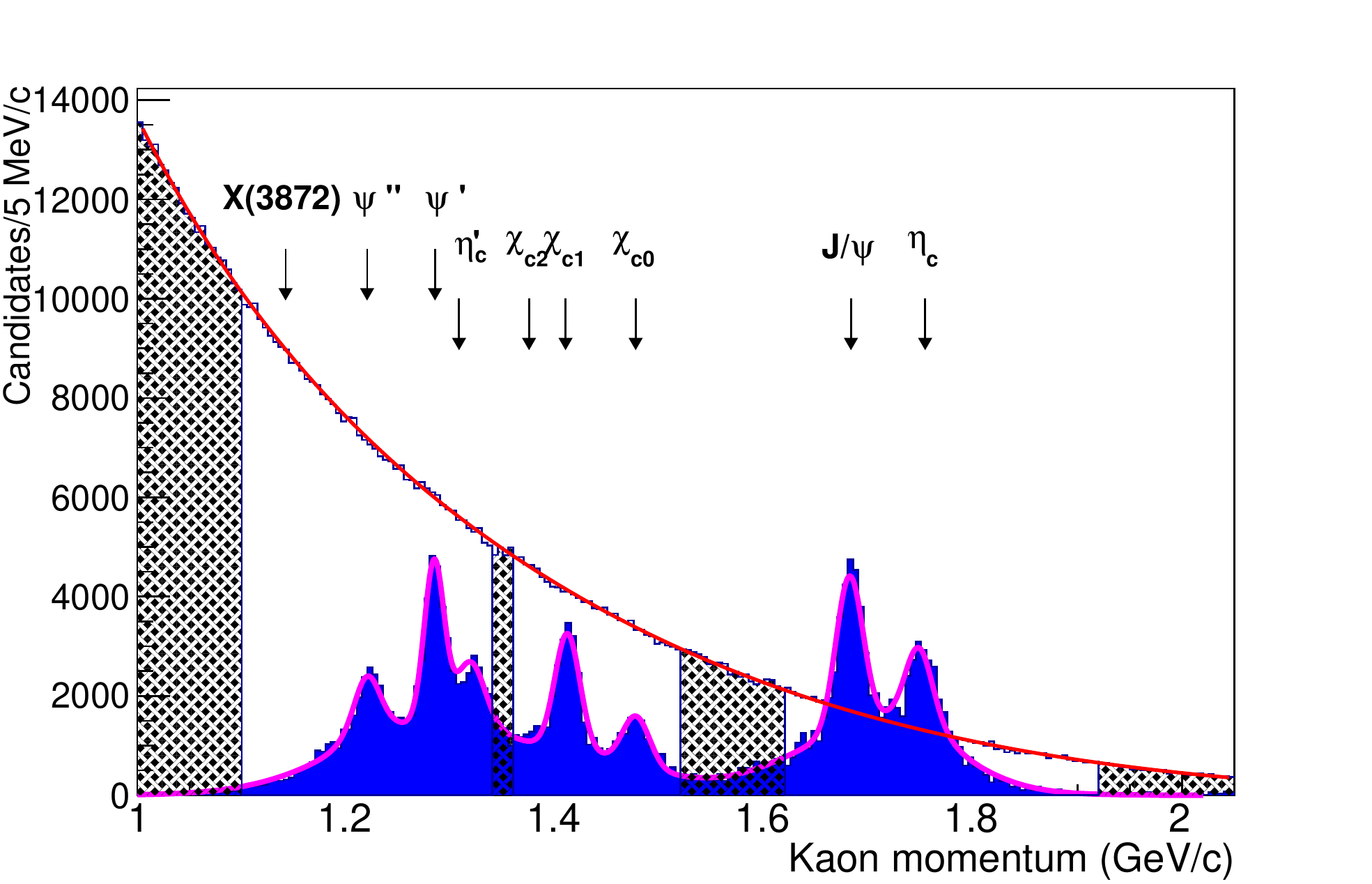,width=.49\textwidth}
\caption{The \Kch\ momentum spectrum for simulated events where no signal kaons are present. The MC statistics represent 3.5 times \babar~ integrated luminosity. The hatched areas correspond to the 
zones used to fit the polynomial background. The filled blue histogram is the signal-only \Kch\ momentum spectrum in simulated events. The purple line represents the fit to this distribution.}
\label{fig:nosignal_polfit}
\end{figure}

To analyze the kaon momentum spectrum we first determine the background shape and second perform a fit to the background-subtracted spectrum.
The shape of the background spectrum is determined by interpolating through regions where no signal is expected, below 1.1 and above 1.9 GeV/$c$. Because the use of only these two regions leads to large uncertainty in the background parameters, we add data points in the two regions 1.34--1.36 GeV/$c$ and 1.53--1.57 GeV/$c$, where there is no peak, as indicated on Fig.~\ref{fig:nosignal_polfit}.

\begin{figure}
\epsfig{figure=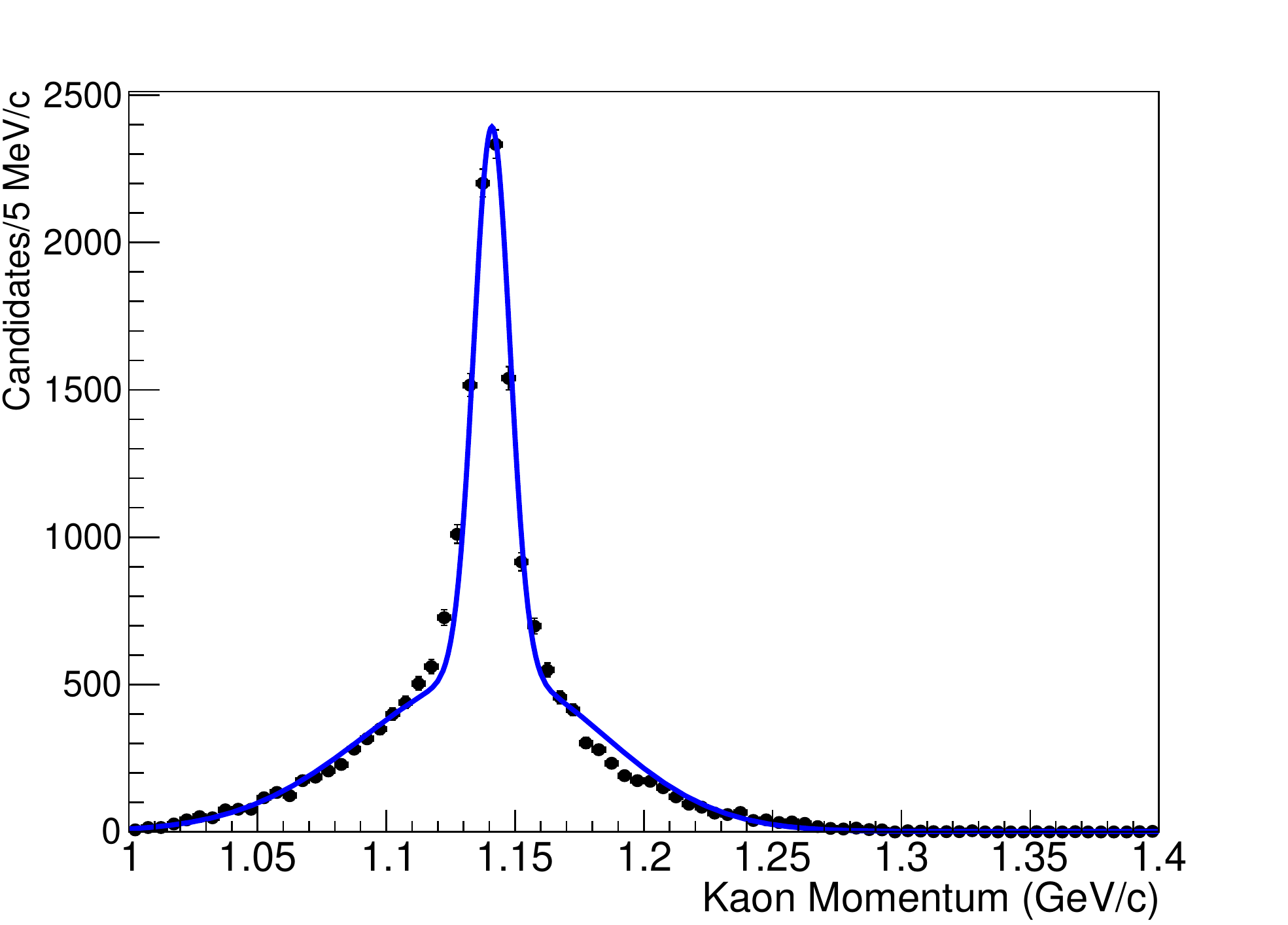,width=.49\textwidth}
\caption{Fit to the signal-only \Kch\ momentum spectrum in $X(3872)$ simulated events.}
\label{fig:x3872_signalonly}
\end{figure}

Figure~\ref{fig:nosignal_polfit} also shows the fit to the simulated signal \Kch\ momentum spectrum for all charmonia peaks in the simulation. A good description is obtained when using, for  each peak, a narrow Gaussian, whose width depends on momentum varying from 13 MeV/$c$ for the \jpsi to 9 MeV/$c$ for the \psip, and a two-piece Gaussian, 100 MeV/$c$ wide on the left  and 60 MeV/$c$ wide on the right. A similar fit is performed for the $X(3872)$ with a dedicated Monte Carlo sample (Fig.~\ref{fig:x3872_signalonly}). The narrow Gaussian width is measured to be 7 MeV/$c$ and the wide Gaussian tails are 47 MeV/$c$ on each side. All parameters describing the shapes of the signal peaks are  fixed to these  values in the fit to data. The wide Gaussian is associated with candidates where the $B$-tag has a reconstructed \mes in the signal region but is not built with the correct  set of $B$ decay products and, therefore, provides an incorrect boost.  The presence of $D$ mesons in \psipp~ or $X(3872)$ leads to a higher background under the $B$ peak, leading to a large wide gaussian component, and  a higher efficiency for the $X(3872)$: the MC efficiency is found to be (48$\pm$2)\% and (25$\pm$0.7)\% for \jpsi  using the low  and high mass training, respectively, (51$\pm$2)\% for \etac, (56$\pm$3)\% for $\chi_{c1}$, (61$\pm$3)\% for \psip, and (77$\pm$2)\% for $X(3872)$.

When using the intermediate points to interpolate the background,  the tails from the \jpsi and \etacp\ peaks extending into these intermediate regions are subtracted using the simulation with the known branching fractions~\cite{PDG}. 
The fit function is a product of fifth-order Chebyshev polynomials and an exponential function. 

Small deviations are observed in the simulation between the background kaon momentum distribution and the fit function~\cite{supplement}. These defects in background shape do not affect the visibility of narrow peaks, such as that of the $X(3872)$ since the expected width of 7 MeV/$c$  is much smaller than the $\sim50$ MeV/$c$ typical width of the local deviations. The observed residuals in the 1.1 to 1.2 GeV/$c$ region are  corrected for, and the resulting uncertainty taken into account.

\begin{figure}
\epsfig{figure=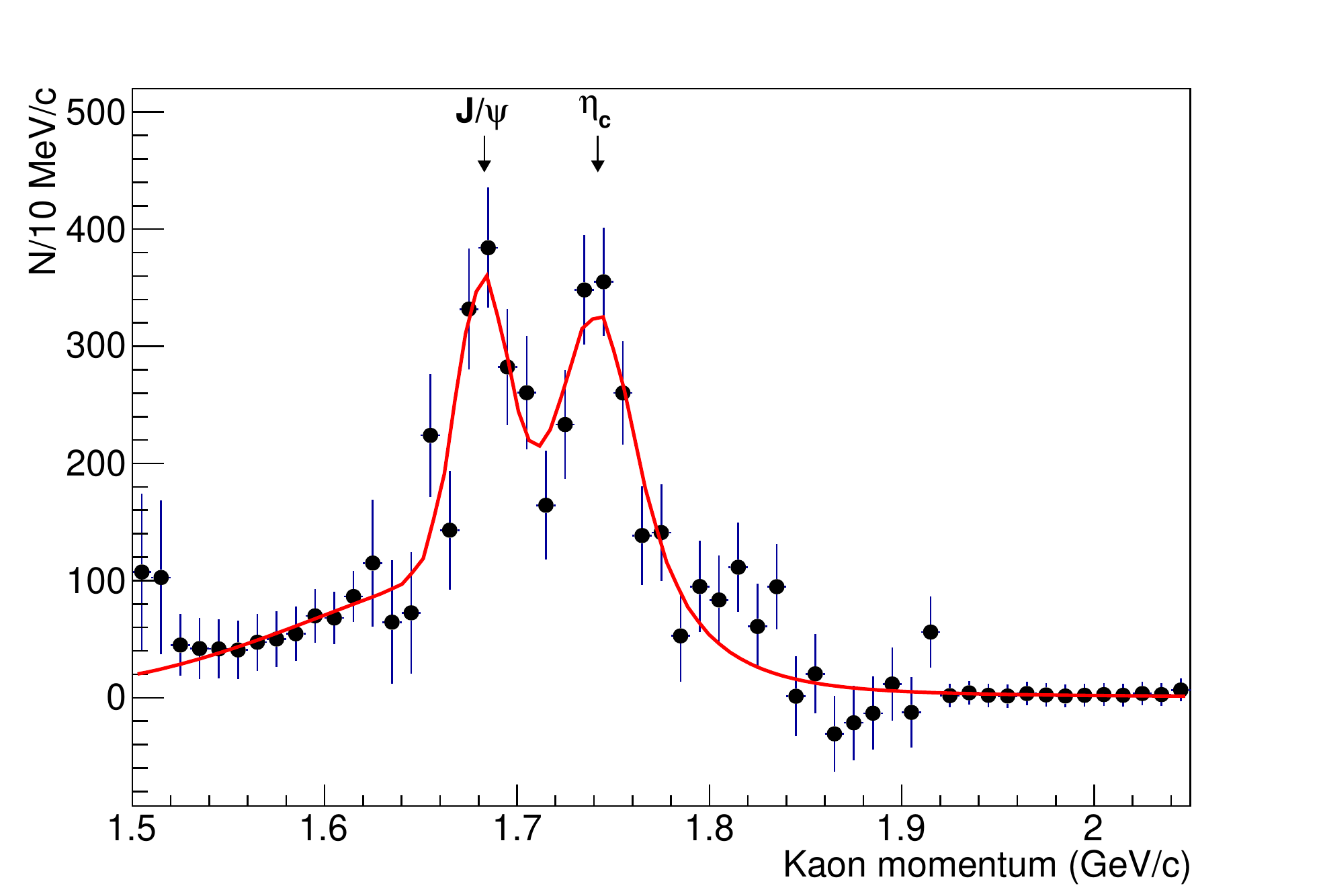,width=.49\textwidth}
\caption{The background-subtracted kaon momentum spectrum for data  in the  $J/\psi -\eta_c$ region with fit result superimposed.}
\label{fig:psisignal_data}
\end{figure}

The kaon spectrum between 1.5 and 2 GeV/$c$ is expected to exhibit two peaks, one at $p_k=1.684$ GeV/$c$ corresponding to the \jpsi  and a second at $p_k=1.754$ GeV/$c$ for the \etac.  The SuperNN is trained in the \jpsi--\etac\  region and the SuperNN output is required to be $>0.85$ with a $B$ purity larger than 0.08. 
A fit to the background-subtracted spectrum is performed with the  two  signal functions determined above, the only free parameters being the charmonia yields. Fig.~\ref{fig:psisignal_data}~displays the results, with the yields: $N_{\jpsi}=2364\pm189$ and $N_{\eta_c}= 2259\pm188$. The statistical precision is  8\%, a factor of about two improvement compared to Ref.~\cite{charmo2005}.

The branching fraction  ${\cal B}(B^\pm\to K^\pm\eta_c)$ is computed using the world average ${\cal B}(B^\pm\to K^\pm\jpsi )$~\cite{PDG}~and the ratio of the yields quoted above, to obtain:
\begin{align}
 {\cal B}&(B^\pm\to K^\pm\eta_c)=\nonumber\\&(0.96\pm0.12({\rm stat})\pm0.06({\rm syst})\pm0.03(\hbox{``ref''}))\times 10^{-3},\nonumber
\end{align}
 where the systematic uncertainty is detailed in Table~\ref{systematics_tab}, and {\rm ``ref''} refers to the uncertainty in ${\cal B}$(\bch$\to K^\pm\jpsi)$ \cite{PDG}. 
This result agrees with the world average, $ (1.09 \pm0.09)\times 10^{-3}$ ~\cite{PDG}. As a cross-check, ${\cal B}(B^\pm\to K^\pm\jpsi)$ is also extracted from the ratio of observed $J/\psi$ events obtained in data and simulation:  ${\cal B}(B^\pm\to K^\pm\jpsi)=(1.09\pm0.09({\rm stat})\pm0.06({\rm syst}))\times10^{-3}$, in agreement with the world average.

\begin{figure}
\epsfig{figure=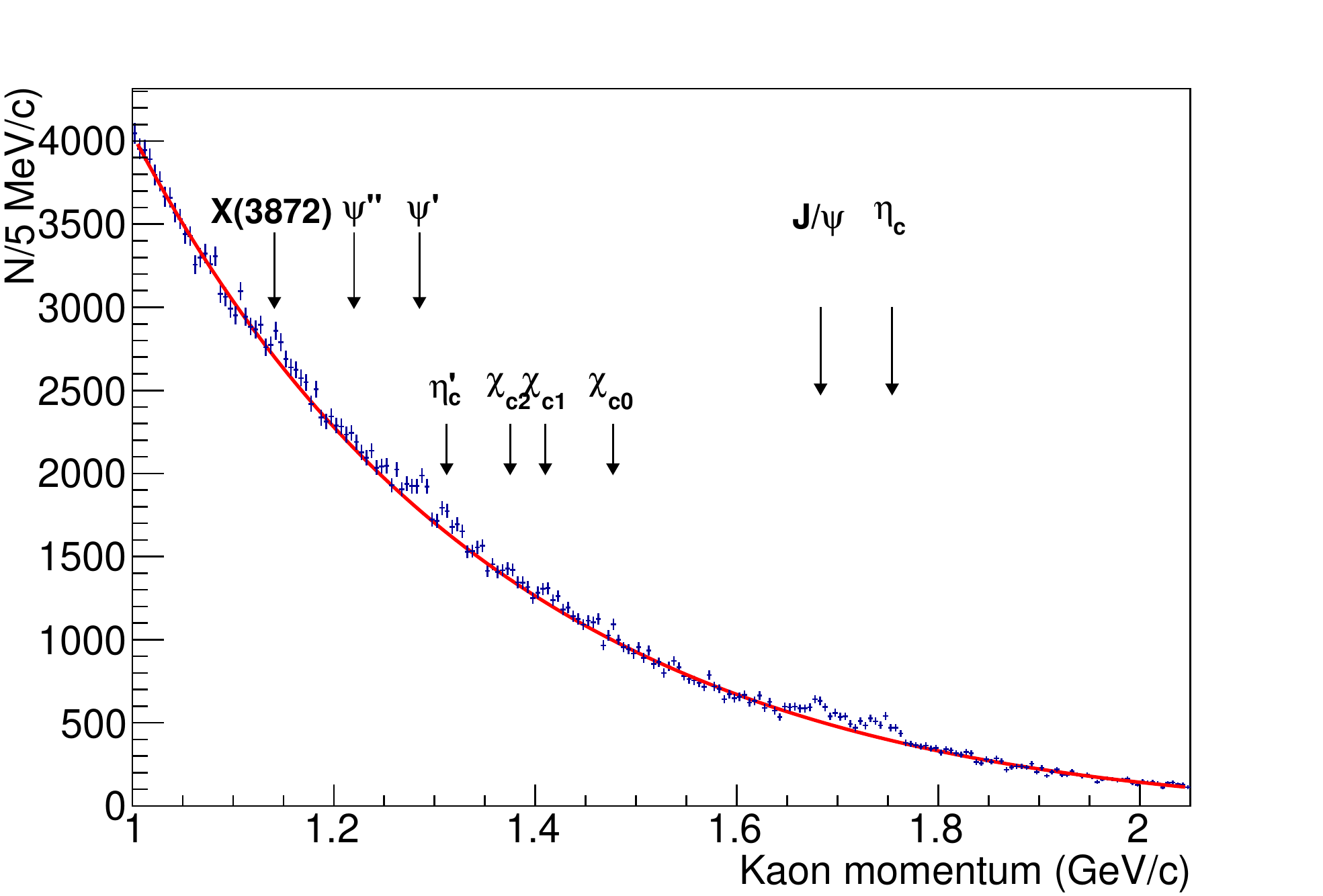,width=.49\textwidth}
\caption{The kaon momentum spectrum after applying final selection criteria and before background subtraction. The red line is the interpolated function describing the background shape. The arrows indicate the values at which a signal for each resonance is expected.}
\label{fig:before_background_subtraction}
\end{figure}

\begin{figure}
\epsfig{figure=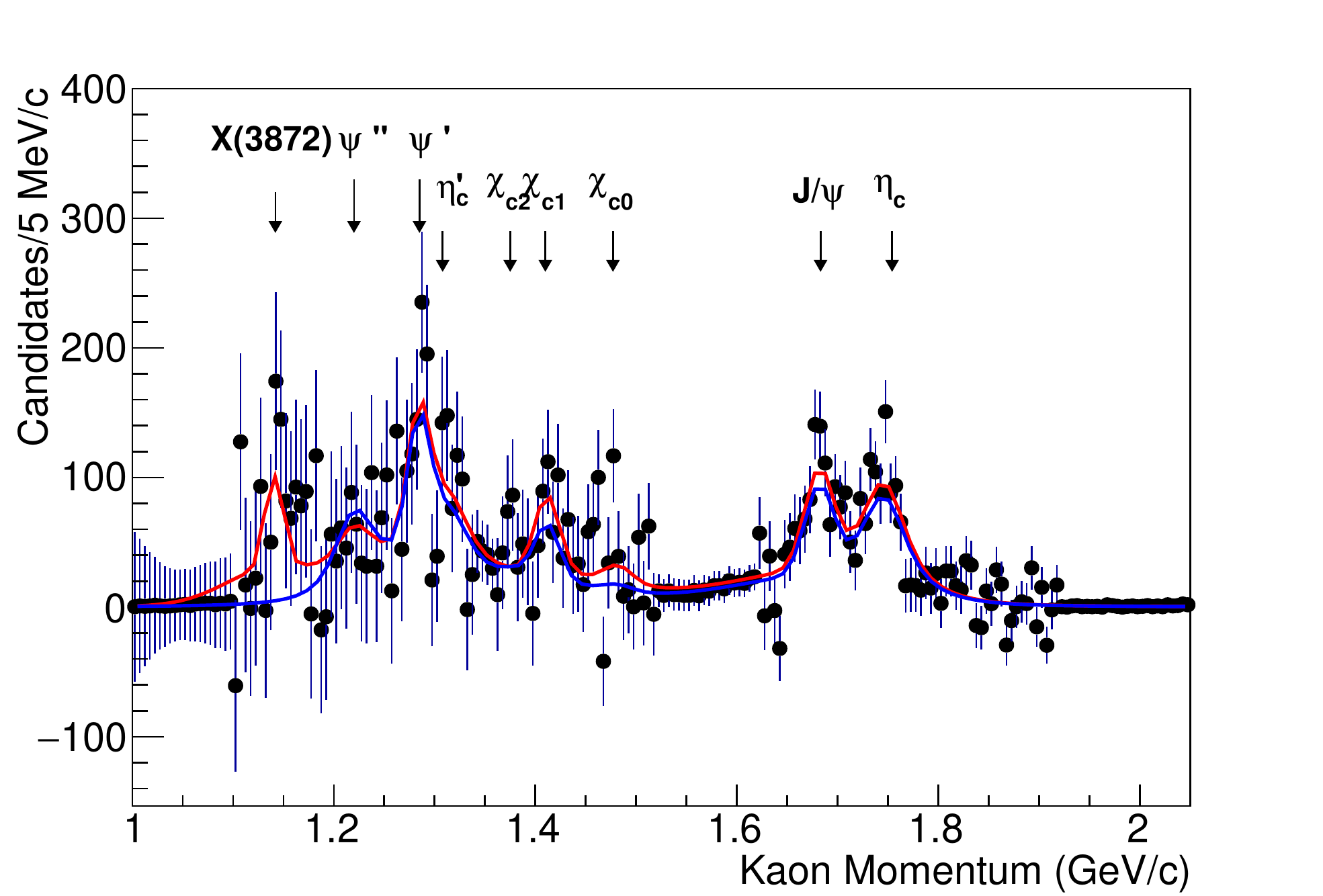,width=.49\textwidth}
\caption{The background-subtracted kaon momentum spectrum between 1 and 2.05 GeV/$c$. The fit function (red) includes signal peaks for nine particles, indicated by the arrows. The fit function where the X(3872) yield is forced to 0 is drawn in blue.
}

\label{unblind_gg12}
\end{figure}

The higher-mass region was blinded during the  initial part  of the analysis. Here, the SuperNN is trained in the \psip\ region and the SuperNN output is required to be $>$0.6 with a $B$ purity larger than  0.10.  The $p_K$ spectrum is fitted using the same procedure as above. The background shape is determined using a fit
 to the signal-free region after correction for the small  residual signal in that region estimated from MC simulation. 
The kaon  spectrum before (after) background subtraction is displayed in Fig.~\ref{fig:before_background_subtraction} (Fig.~\ref{unblind_gg12}).

 The fit to the background-subtracted signal spectrum (Fig.~\ref{unblind_gg12}) is a sum of nine signal-peak functions corresponding to the $X(3872)$, \psipp, $\psi^\prime$, \etacp, $\chi_{c2}$, $\chi_{c1}$, $\chi_{c0}$, \jpsi,  and \etac. The peak locations are taken  from Ref.~\cite{PDG} and the widths  from fits to MC signal samples and include both detector resolution and the natural width of each resonance. The peak labelled \chione refers to both \chione and \hc\ since these two states cannot be distinguished from each other in this analysis.   A binned maximum
 likelihood fit is performed, with the nine charmonium yields as free parameters.  
 Table \ref{final_fits} contains the fit results. Signal peaks are visible for \etac, \jpsi, $\chi_{c1}$, $\psi^\prime$~\cite{supplement},  and $X(3872)$.  A separate fit in which the $X(3872)$ signal is forced to 0 has a \chisq larger than that of the nominal fit by 11.1 units, which reduces to 9.0 when accounting for the uncertainty in the background shape in the 1.1 to 1.2 GeV/$c$ region. Thus, there is 3$\sigma$ evidence of  the decay \bch\to\Kch $X(3872)$, detected  for the first time using this recoil technique.

Systematic uncertainties mainly stem from the imperfect description of the data by the simulation, and are computed for the five particles having  significance $>2\sigma$. An extra uncertainty is added for the X(3872) for the  limited knowledge of its decay modes.
\begin{itemize}
\item {\it Peak position.}
A deviation from the known peak position can induce an uncertainty in the number of events from the fit integral, estimated at 1\%. 
\item {\it Signal shape.} Four  parameters are used to describe the signal shape: the main narrow width of the signal peak, the widths of the left-hand side and right-hand side Gaussian tails, and the  fraction under the narrow Gaussian. The  uncertainty resulting from the uncertainty in the signal shape is estimated using the fit to the  simulation sample containing only true kaons from two-body \bch\ decays by comparing the fit results with the true numbers of events.
 When the resonance has a non-negligible natural width, as for the \etac, the uncertainty in this width is  included. 
\item {\it Background subtraction.} The statistical uncertainty of the background fit is propagated, including correlations, into the  statistical uncertainty and is not  a systematic uncertainty. The systematic uncertainties stem from different background parametrizations and from the correction due to the signal subtraction in the 1.1 to 1.2 GeV/$c$ region. This latter uncertainty is determined as the change to the $X(3872)$ yield introduced by a one-sigma deviation of the correction function.
\item {\it Efficiency determination.} Uncertainties in detection efficiency arise in the kaon reconstruction and particle identification, and in the SuperNN-based selection. These uncertainties cancel to a good approximation in the ratios of the branching fractions of all resonances to the \jpsi. 
\item {\it X(3872) decay model.} The signal shape is not the same for $DD$ and $\jpsi$X decays and this effect induces a small change in the signal yield in the fit. Varying the ratio between these two types of decays leads to a 5\% additional uncertainty.
\end{itemize}
Table~\ref{systematics_tab} summarizes the various systematic uncertainties, and  Table ~\ref{final_fits}  summarizes the branching fraction results.
\begin{table}[t]
\begin{center}
\begin{tabular}{|l|c|c|c|c|c|} \hline
Uncertainty source  &  \etac&\chione&\etacp&\psip&$X(3872)$  \\
\hline\hline
$K$ identification &  1 &2&2&2  & 5 \\
Decay model& -  &-&1&-& 5  \\
Efficiency  &0&2&2&2&5\\
$p_K$ : peak position &  2&2&8&2& 2 \\
$p_K$ : signal narrow width &  1&1&1&1&1 \\
$p_K$ : signal wide width &  5&5&5 &5  & 5 \\
$p_K$ : narrow width fraction  &  2&2&2&2  & 2 \\
$p_K$ : background shape&-&13&12&13 &13    \\
Decay width&1&-&1&-&-\\
Correction in signal-free regions&-&-&-&-&4\\
\hline
Total &  6&14.5&15.1&14.6  & 16.3 \\
\hline
\end{tabular}
\caption{Summary of relative systematic uncertainties (in \%) for the \etac, \chione, \etacp, \psip, and $X(3872)$ branching fractions, relative to ${\cal B}(B^\pm\to\jpsi K^\pm)$.}
\label{systematics_tab}
\end{center}
\end{table}

\begin{table}[t]
\begin{center}
\begin{tabular}{|c|c|c|c|} \hline
Particle &  Yield  &${\cal B}$(10$^{-4}$)&N$_{\sigma}$\\
\hline
\jpsi&\quad\hskip3pt 2364$\pm$189 \quad\quad &\quad 10.1$\pm$0.29 (Ref.~ \cite{PDG}) \quad&10.4 \\
\etac&2259$\pm$188&9.6$\pm$1.2(stat)$\pm$0.6(syst)&9.3\\
\hline
\chizero &     287$\pm$181   & 2.0$\pm$1.3(stat)$\pm$0.3 (syst)&1.6      \\
\chione	 &    1035$\pm$193    &4.0$\pm$0.8(stat)$\pm$0.6(syst)&2.2       \\
\chitwo	 &   200$\pm$164      & $<$2.0&1.2       \\
\etacp& 527$\pm$271& 3.5$\pm$1.7(stat)$\pm$0.5(syst)&2.3\\
\psip	 &  1278$\pm$285&4.6$\pm$1(stat)$\pm$0.7(syst)&3.1      \\
\psipp& 497$\pm$308 &3.2$\pm$2.0(stat)$\pm$0.5(syst)&1.2 \\
$X(3872)$& 992$\pm$285 & 2.1$\pm$0.6(stat)$\pm$0.3(syst)&3.0 \\
\hline
\end{tabular}
\caption{Results from fits to the kaon momentum spectrum. ${\cal B}$ stands for the branching fraction for  \bch\to X$_{c\bar c}$\Kch. An additional 3\% uncertainty must be added to these results, reflecting the present knowledge of the reference ${\cal B}$($B^+$\to\jpsi\Kp). The significance of each peak refers to the \chisq increase of the fit when removing each resonance in turn.} 
\label{final_fits}
\end{center}
\end{table}

 The number of $X(3872)$ events is converted into an absolute branching fraction using the number of observed \jpsi events, its absolute branching fraction, and the relative efficiency ratio, with the result:
${\cal B}(\Bp\to X(3872)\Kp )= (2.1\pm 0.6({\rm stat})\pm0.3({\rm syst})\pm0.1(\hbox{``ref''}))\times 10^{-4}$.
 Using the measured product branching fraction  ${\cal B}(\B^+\to X(3872)K^+)\times {\cal B}(X(3872)\to J/\psi\pi^+\pi^-) = (8.6\pm0.8)\times10^{-6}$~\cite{PDG}, this translates into ${\cal B}(X(3872)\to J/\psi\pi^+\pi^-)= (4.1\pm1.3)\%$. From this, an upper limit on the partial width $\Gamma(X(3872)\to J/\psi\pi^+\pi^-)$ can be set in the  100 keV range, using 3 MeV as an upper limit for the $X(3872)$ total width, as measured in its $DD$ decay channel \cite{babardd, belledd}.
 Our measurement  therefore suggests that the $X(3872)$ has a significant molecular component.
 
We report an  update to our first analysis~\cite{charmo2005} with the full \babar\ statistics. Two new features are introduced: the inclusion of all $B$ candidates has led to an increase of efficiency and a better separation between signal and tag sides of an event; the fit to a polynomial background in regions where no signal is present reduces the statistical and systematic uncertainties related to the background subtraction. We obtain the following results:
\begin{equation*}
\begin{split}
  {\cal B} &(B^+\to\eta_c K^+)=\\&(0.96\pm0.12({\rm stat})\pm0.06({\rm syst})\pm0.03({\rm ref}))\times 10^{-3},\\
 {\cal B} &(B^+\to X(3872)K^+)=\\&(2.1\pm0.6({\rm stat})\pm0.3({\rm syst})\pm0.1({\rm ref}))\times 10^{-4},\\
 {\cal B} &(X(3872)\to J/\psi\pi^+\pi^-)= (4.1\pm1.3)\%.
\end{split}
\end{equation*}
This result will certainly contribute to the determination of the complex nature of the X(3872) particle.

We are grateful for the 
extraordinary contributions of our \pep2\ colleagues in
achieving the excellent luminosity and machine conditions
that have made this work possible.
The success of this project also relies critically on the 
expertise and dedication of the computing organizations that 
support \babar.
The collaborating institutions wish to thank 
SLAC for its support and the kind hospitality extended to them. 
This work is supported by the
US Department of Energy
and National Science Foundation, the
Natural Sciences and Engineering Research Council (Canada),
Institute of High Energy Physics (China), the
Commissariat \`a l'Energie Atomique and
Institut National de Physique Nucl\'eaire et de Physique des Particules
(France), the
Bundesministerium f\"ur Bildung und Forschung and
Deutsche Forschungsgemeinschaft
(Germany), the
Istituto Nazionale di Fisica Nucleare (Italy),
the Foundation for Fundamental Research on Matter (The Netherlands),
the Research Council of Norway, the
Ministry of Science and Technology of the Russian Federation, and the
Particle Physics and Astronomy Research Council (United Kingdom). 
Individuals have received support from 
CONACyT (Mexico),
the A. P. Sloan Foundation, 
the Research Corporation,
and the Alexander von Humboldt Foundation.


\begin{thebibliography}{9}

\bibitem{quigg}

 C.~Quigg, FERMILAB-Conf-04/033-T, and hep-ph/0403187, and references therein.
\bibitem{babar1}
 B.~Aubert et al. ({\babar} Collaboration), Phys. Rev. D {\bf 67}, 032002 (2003).
                                                                                
\bibitem{belle2} 

S.~K.~Choi et al. (Belle Collaboration),  Phys. Rev. Lett. {\bf 89}, 102001 (2002).
\bibitem{bellex38}
S.~K.~Choi et al. (Belle Collaboration), Phys. Rev. D {\bf 84}, 052004 (2011).
\bibitem{babarx38}
 B.~Aubert et al. ({\babar} Collaboration), Phys. Rev. D {\bf 77}, 111101 (2008).
\bibitem{maiani}
L.~Maiani et al., Phys. Rev. D {\bf 71}, 014028 (2005).
\bibitem{maiani2}
L.~Maiani et al., Phys. Rev. Lett. {\bf 99}, 182003 (2007).

\bibitem{braaten}
 E.~Braaten et al.,  Phys. Rev. D {\bf 72}, 054022 (2005).
\bibitem{barnes} 
T.~Barnes, S.~Godfrey, Phys. Rev. D {\bf 69}, 054008 (2004).
\bibitem{ortega} 
P.~G.~Ortega, E.~R.~Arriola, Chin.Phys. {\bf C43} 12, 124107 (2019).
\bibitem{swanson}
 E.~Swanson, Phys. Lett. B {\bf 588}, 189 (2004).
\bibitem{charmo2005}
 B.~Aubert et al. ({\babar} Collaboration), Phys. Rev. Lett. {\bf 96}, 052002 (2006).
\bibitem{belle:similar}
Y.~Kato et al. (Belle Collaboration), Phys. Rev. D {\bf 97},  012005 (2018).

\bibitem{babarlumi}
J.P. Lees et al. ({\babar} Collaboration), Nucl. Instr. and Methods A {\bf 726}, 203 (2013).
\bibitem{babardet1}
B. Aubert et al. ({\babar} Collaboration), Nucl. Instum. and Meth. A \textbf{479}, 1 (2002).
\bibitem{babardet2}
B.~Aubert et al. ({\babar} Collaboration),  Nucl.~Instr.~Meth. A {\bf 729},  615 (2013).

\bibitem{fullreco}
 B.~Aubert et al. ({\babar} Collaboration), Phys. Rev. Lett. {\bf 109}, 101802 (2012). 
 
\bibitem{crystalball}

T.~Skwarnicki et al., DESY-F31-86-02.

 \bibitem{argus}

  H.~Albrecht et al. (ARGUS Collaboration), 

Phys. Lett. {\bf B318}, 397 (1993).                                                     
\bibitem{supplement}
See Supplemental Material at [URL will be inserted by publisher] for the performance of the topological NN (Fig. 1), for the distribution of the residues to the fit in MC, as function of  kaon momentum (Fig. 2), for the analysis results, when optimized in the \psip-\etacp~region, as function of momentum (Fig.3), and of recoil mass (Fig. 4).

\bibitem{PDG}
M. Tanabashi et al. (Particle Data Group), Phys. Rev. D {\bf 98}, 030001 (2018) and 2019 update. 

\bibitem{babardd}
B. Aubert et al. ({\babar} Collaboration), Phys. Rev. D {\bf 77}, 011102 (2008).
\bibitem{belledd}
T.~Aushev et al. (Belle Collaboration), Phys. Rev. D {\bf 81}, 031103 (2010).
\end{thebibliography}
\end{document}